# Multi-Objective Optimisation with Oscillatory Dynamics in Spontaneous and Decision Spiking Neural Networks

Divyansh Sethi[1], Muhammad Faraz[1], and KongFatt Wong-Lin[1]
[1]Intelligent Systems Research Centre, School of Computing, Engineering and Intelligent Systems,
Ulster University, Magee Campus,
Derry~Londonderry, Northern Ireland, UK
Sethi-D@ulster.ac.uk, Faraz-M1@ulster.ac.uk, k.wong-lin@ulster.ac.uk

*Abstract*— **Spiking neural networks (SNNs) can be used for implementing cost-efficient artificial intelligence computing or mechanistic modelling of experimentally observed neural data. In the latter, fitting neural data with recurrent SNNs (RSNNs) remains a challenge. Importantly, given that neuronal network oscillations are known to play important roles in neural functions, fitting specific RSNN oscillation frequencies with neural firing rates has yet to be fully explored. In this work, we extended our previous application of genetic algorithm (GA), specifically non-dominated sorting GA (NSGA-III), on sensitive Izhikevich neuron-based RSNNs by optimising their connectivity parameters to target emergent neuronal (sub)population firing rates and network oscillation frequencies. We evaluated this, via RMSEs on a Pareto frontier, on spontaneously active simulated RSNN model and low-activation brain organoid, followed by a simulated RSNN model with transient decision dynamics. In all cases, the models comprised spontaneously firing cortical excitatory and inhibitory neurons. We showed that NSGA-III could readily optimise for multiple network firing rates and dominant network oscillation frequencies, and for the decision-making model, for activity patterns in different time epochs. Notably, dominant oscillation frequencies were found to be more parameter sensitive, but firing rates were more robustly met. We also identified low-activity regime for decision-making. Overall, we have successfully demonstrated the implementation of multi-objective GA optimisation on RSNNs' and brain organoid's neural firing rates and oscillations.**

## I. INTRODUCTION

Spiking neural networks (SNNs) mimic certain functions of real biological brains [1][2], encoding information in neuronal spike times while using less computational resources than traditional neural network models [3][4]. SNNs have been used for deep learning. data streams like audio and video processing for classification, and for neuromorphic computing systems for low-cost edge analytics [5][6][7].

SNNs have also been used to model the biological brain [8][9][10]. Based on spike times, SNNs can also encode information in its firing frequencies (or firing rates), and certain cognitive functions, such as decision-making via pooling from neuronal population [8][9]. The discretised neuronal spiking activities could also lead to emergent neuronal network oscillations [9]. In particular, oscillatory dynamics, measurable by invasive (e.g. multielectrode array MEA, local field potential LFP, and intracranial electroencephalography iEEG) or non-invasive (e.g. scalp EEG, magnetoencephalography, or functional magnetic resonance imaging fMRI) neural recordings [11][12][13], have been known to be associated with specific cognitive and brain states [14][15][16].

There are various ways to explicitly model spiking neurons, depending on the needs [9][10]. A good balance between biological realism of spiking characteristics and computational efficiency is the Izhikevich neuronal model by operating in the vicinity of dynamic bifurcations [10][17][18][19][20][21]. Functional applications of the model include classification [22], emergent/self-organisation [23][24], and associative memory [25]. However, it is also this reason, the modelling formalism is sensitive in terms of its neuronal model parameters. Adding to this challenge, is the well-known difficulty of learning recurrent neural network (RNN) models' parameters [26]. Hence, learning model parameters for Izhikevich neuron-based RNNs can become particularly challenging.

This falls under the study of optimisation which involves algorithmically search for some optimal solution(s) with respect to the model parameter(s) and some specified goal(s) (i.e. some objective mathematical function(s)) [27]. Popular optimisation techniques include maximum likelihood estimation, gradient descent and genetic algorithm (GA). In particular, GA, a class of evolutionary algorithms, which is inspired by biological evolution theory, is a heuristic, stochastic, randomised search optimisation technique and it is relatively simple to describe and implement [28][29][30]. GA requires no *a priori* knowledge about what it is trying to optimise as domain specific knowledge is contained in the fitness function and the genetic operators defined for the problem.

In our previous investigations [31][32], we had applied GA to Izhikevich neuron-based RNNs. Specifically, our more recent work [32] had applied multi-objective GA, particularly non-dominated sorting GA (NSGA-III) [33], to optimise connectivity parameters for target firing rates averaged over individual neuronal population and (simulated) time. Other studies have investigated synchronised or spontaneous network oscillation network frequency [23][34] but not simultaneously with firing rates.

Together, these previous studies did not explore the possibility of multi-objective optimisation (MOO) that comprise both neural firing rates and network oscillation frequencies. Further, the evaluations of NSGA-III on brain organoid (for organoid computing) and decision-making network with transient firing rates and oscillatory dynamics are currently lacking. These cases are particularly challenging, given possible low neuronal firing rates in brain organoids [35] and non-stationary nonlinear neural dynamics in decision-making [36][37].

In this work, we address the above knowledge gap by employing GA on three different cases: (1) spontaneous activities of a canonical cortical recurrent SNN (RSNN) model [17]; (2) a RSNN model that replicates spontaneous activities of an actual biological brain organoid sample [38]; and (3) low but transient activities of a binary decision-making SNN with biophysically realistic synapses. In each of the three cases, excitatory and inhibitory neuronal model of the Izhikevich formalism was used [17] and the dominant network oscillation frequency identified. NSGA-III was used for multi-objective optimisation on model connectivity parameters with respect to minimising the root-mean-square errors (RMSEs) of neuronal population firing rates and dominant network oscillation frequencies from their target values.

Building on this framework, the work makes key contribution of evaluating a unified NSGA-III optimisation pipeline that jointly fits firing-rate, oscillatory, and behavioural targets across canonical cortical, organoid-inspired and decision-making RSNN models.

## II. SPIKING NEURAL NETWORK MODELLING

### A. Izhikevich neuronal model

The Izhikevich neuronal model is a phenomenological model that mimics biologically realistic spiking patterns without involving the modelling of a variety of ion channel currents [1]. The (trans)membrane potential v of the model can be described by the coupled differential equations [17]:

$$\frac{dv}{dt} = 0.04\,v^2 + 5\,v + 140 - w + I \qquad (1)$$

$$\frac{dw}{dt} = a(b\,v - w) \qquad (2)$$

where $w$ is some recovery variable coupled to $v$, $I$ is the total afferent/input current, and $a$ and $b$ are model parameters that partially determine the spiking characteristics. The model parameter $a$ describes the rate of decay for the neuronal membrane potential $w$ while parameter $b$ represents $w$'s sensitivity to $v$.

When $v$ passes some prescribed peak (set at 30 mV), this results in a neuronal firing or a spike of activity. Upon firing, $v$ is reset to some level $c$ (-65 mV), another model parameter, while the recovery variable $w$ is simultaneously raised by a value of $d$. Both $c$ and $d$ are two additional model parameters, randomised across neurons with values of -65 mV and 2, respectively, which brings to a total of 4 neuronal parameters. These 4 parameters can alter the spiking behavior of a neuron.

### B. Synapses and network

*Case 1 (spontaneous canonical cortical RSNN activities):* The SNN model consists of 800 excitatory (regular spiking) neurons and 200 inhibitory (fast-spiking) neurons, based on the observed ratio of excitatory to inhibitory neurons in the mammalian cortex to be around 4:1. Following previous study [32], we used instantaneous current-based synapses [17], i.e. whenever a presynaptic neuron fire a connected postsynaptic neuron will receive an instantaneous pulse-like increase (or decrease) by some value in the postsynaptic current through $I$ in Eqn. (1), if the synapses are excitatory (inhibitory).

The range of the synaptic weights can be adjusted with the parameters $g_e$ for excitatory neurons and $g_i$ for inhibitory neurons. The connections between all neurons are represented by some matrix $S$. When investigating network sparsity, some connections in $S$ are set to 0 to mimic sparsity in the network, with fraction of connections, $f$ (= total number of connections in any considered model divided by total possible connections for an all-to-all connectivity model), randomly selected between 0 and 1.

The 5 model parameters to be optimised were $ge$, $gi$, $f$, and thalamic input to excitatory and inhibitory neurons. The 3 objective functions to be minimised were the RMSEs, with respect to target excitatory neuronal population firing rate (time-averaged over 1 second of simulated time) and that of the inhibitory neuronal population, and the dominant network oscillation frequency. The latter is obtained from the highest peak frequency from standard power spectral density (PSD) of the summed input current, i.e. mimicking LFP, of the network.

*Case 2 (low spontaneous activities of brain organoid):* Electrophysiological data of brain organoid sample recorded by FinalSpark's Neuroplatform [38] was used. Further experimental details should be referred to the original experimental study [38]. The sample data used consisted of 13 provided neurons simultaneously recorded continuously over 5 days. Upon inspection via appropriate threshold setting of neuronal membrane potentials, 6 of theses neurons had too few spikes and hence we retained the remaining 7 neurons, specifically neuron numbers 0, 1, 2, 12, 15, 18 and 19.

As a feasibility study, for each neuron, we equally divided the first 60 min of recordings into twelve non-overlapping 300 s time segments to obtain segment-averaged instantaneous neuronal firing rate timecourse, and we observed the low firing rates of individual neurons (Fig. 1). Finally, we averaged over all 7 neurons to obtain the instantaneous neuronal population firing rates (Fig. 1, bottom right). By averaging over time, we found the firing rate to be 0.09 Hz. From the firing rate's PSD, the dominant frequency was 0.195 Hz. These two values were then set as target values for the RSNN model.

Based on the spike waveforms of the data [38], with about 1 and 0.5 ms spike widths, and that the interspike intervals were sufficiently regular, our RSNN model comprised regular spiking and fast-spiking neurons, as in Case 1. Given limited information on specific neuronal type, the model had 5 excitatory regular-spiking neurons and 2 inhibitory fast-spiking neurons, to maintain higher excitatory than inhibitory neuronal numbers with enough inhibitory neurons for network stability. With only 7 neurons, we considered search for model parameters associated with connection weights and stimulus inputs (see below). Only the model's firing rate averaged over the 7 simulated neurons were considered for optimisation. Due to lack of information of the organoid's synaptic activities, we derived the PSD's dominant frequency from the instantaneous neuronal population firing rates. Here, we used GA to optimise for the neuronal population's firing rates and associated dominant oscillation frequency, i.e. two objectives.

*Case 3 (non-stationary nonlinear decision RSNN activities):* A minimal SNN model that performed binary decision-making consisted of 2 excitatory neuronal populations, each with 200 neurons and encoding one of two choices. These two populations could potentially compete in a winner-take-all (WTA) behaviour, mediated by an intermediate inhibitory neuronal population, consisting of 100 neurons. This was based on a similar previous model [37], but here, we incorporated biophysical synaptic models of the AMPA-, NMDA- and GABA(A)-mediated types as in the classic attractor network model [36] to allow for robustness of temporal integration and non-trivial network oscillations [15]. We also sought to identify whether low-activity WTA decision-making is possible in this model.

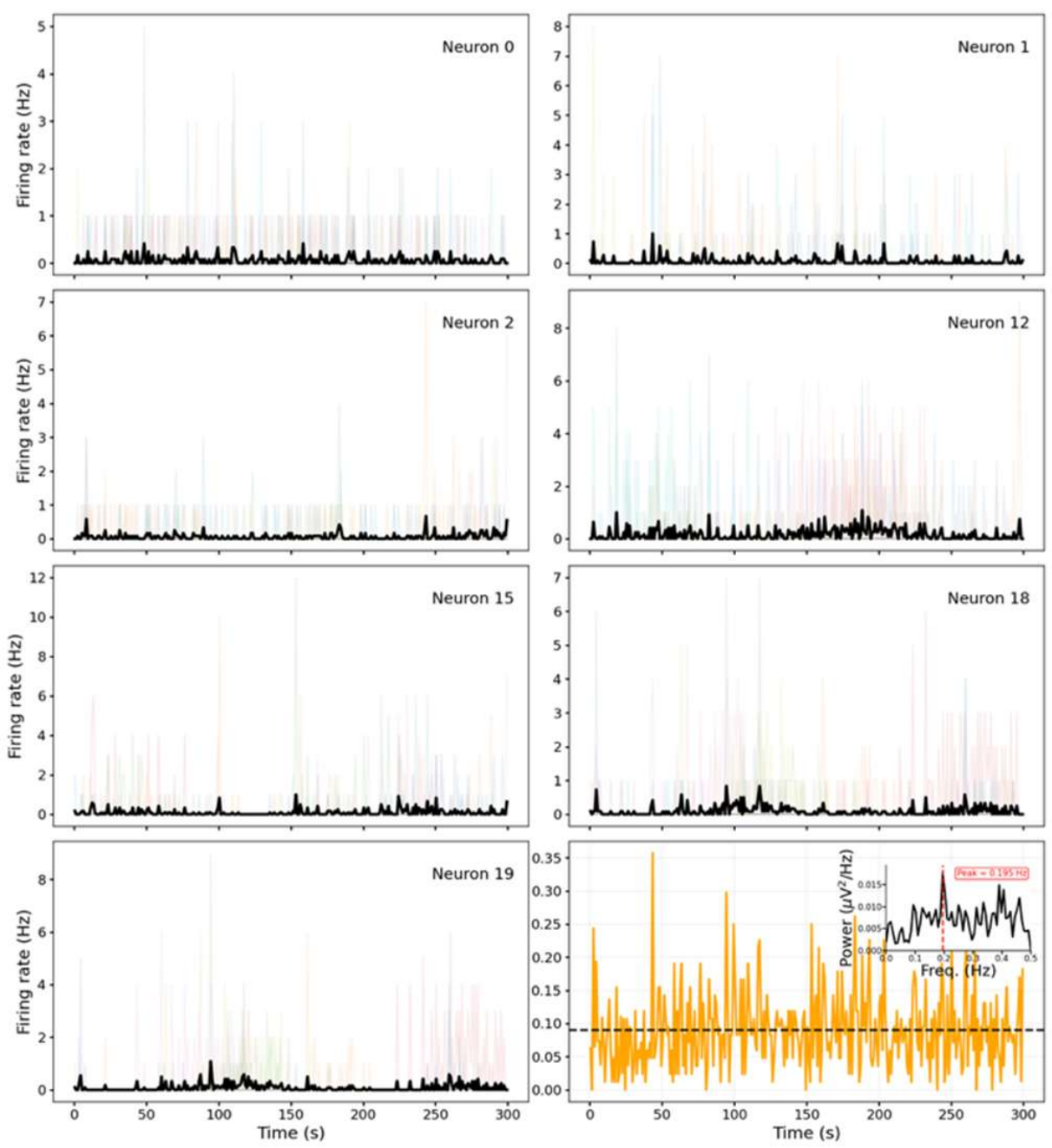

*Fig. 1. Low spontaneous firing rates of 7 organoid neurons. Different light colours: firing rates in each of the 12 300-s segments with 1 s window. Black: firing rates averaged over 12 equal time segments. Bottom right, orange: firing rates averaged over 7 neurons, with time averaged value of 0.09 Hz (dashed). Inset: PSD showed dominant network oscillation frequency at 0.195 Hz (highest peak; red dashed).*

For our study's purpose, unbiased stimuli were presented at 500-1500 ms, in the form of equal input currents to the two excitatory populations while the latter still can still exhibit WTA behaviour, with a competing population attaining higher activities than the other. Biased inputs favoring one of the excitatory populations are generally more trivial and we did not investigate that. Hence, there were 3 key time epochs in a simulated trial: (I) 0-500 ms of spontaneous activities as in Case 1; (II) 500-1500 ms of unbiased stimulus inputs when binary decision had to be made via WTA behaviour; and (III) 1500-2500 ms of post-stimulus duration which might encode (working) memory of the decision made in (II). 30 trials per session were used with about equal chance of either excitatory population winning.

We implemented 2 excitatory ($e$) to inhibitory ($i$) neuronal connection types with weights $W_{e\text{-}i,AMPA}$ and $W_{e\text{-}i,NMDA}$, 1 $i$-to-$e$ connection type with weight $W_{i\text{-}e,GABA}$, 2 self-excitatory ($e$-$e$) connection types with weight $W_{e\text{-}e,AMPA}$ and $W_{e\text{-}e,NMDA}$, and 1 self-inhibitory ($i$-$i$) connection type $W_{i\text{-}i,GABA}$. These, with input current $I_{stim}$, were optimised for the 3 objectives in the following time epochs: (I) activity difference between excitatory populations < 0.2 Hz; (II) activity difference between excitatory populations > 4 Hz and with set peak frequency target of 30 Hz. For simplicity, we did not restrict activities in epoch III and that of inhibitory neuronal population. LFP activities were computed from the summed AMPA-, NMDA- and GABA(A)-mediated synaptic currents.

## III. MULTIPLE-OBJECTIVE GA OPTIMISATION

GA [28][29] starts with an initial, often random, population of chromosomes, each representing a candidate solution encoded as genes. A fitness function evaluates each chromosome, guiding the application of genetic operators. Selection retains the fittest individuals, crossover combines pairs of chromosomes with probability proportional to fitness, and mutation introduces small random changes with low probability to maintain diversity.

In MOO, multiple objective functions (e.g., $G$and $H$) must be optimised simultaneously over a set of parameters $\{p\}$. When objectives conflict, solutions are evaluated using the Pareto frontier, a set of trade-off solutions, where improving one objective would worsen another. Here, the Pareto frontier was obtained by NSGA-III to minimise the objective RMSEs. The overall RMSEs through the RMS of each RMSE was also obtained to identify the lowest set of RMSEs. Specifically, for some specific objective variable $x_1$ (e.g. firing rate or frequency) to be optimised, the RMSE is:

$$RMSE_{x_1} = \frac{1}{N}\sqrt{\sum_{i=1}^{N}(x_{1,i} - x_{1,target})^2} \quad (3)$$

where N is the number of trials per generation. To compute the overall, composite RMSE for all objective variables, we compute $\sqrt{\sum_{j=1}^{M} F_{x_j}^2}$, where M is the total number of objectives to be optimised, and $F_{x_i} = \frac{RMSE_{x_i}}{x_{i,target} + \epsilon}$ is the normalised version for each RMSE. A summary of the methods is shown in Fig. 2.

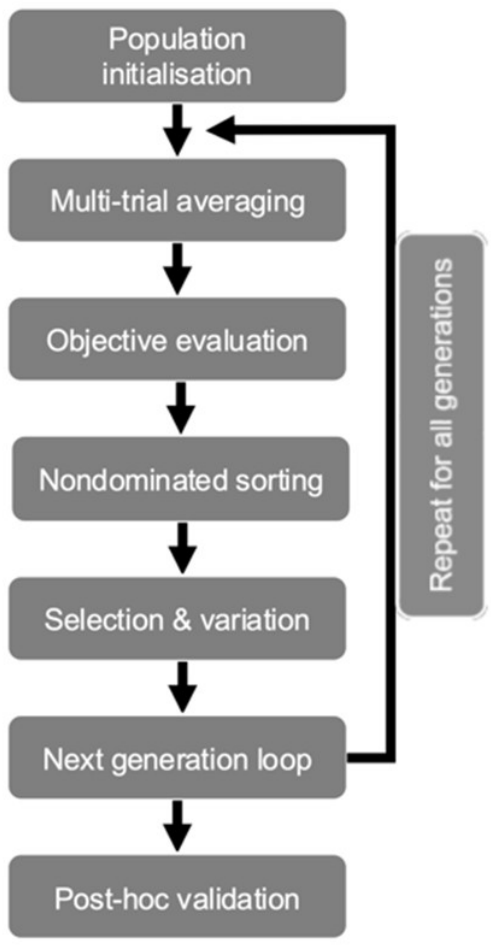

*Fig. 2. Schematic summary of the pipeline used.*

*Software and hardware:* Python 3.10.5 with Python Pandas software library was used for data manipulation and analysis, while Matplotlib and Plotly were used for data visualisation. Simulations and GA (NSGA-III) optimisation used the NI-HPC facility (https://ni-hpc.ac.uk/). Source code and generated data are available at https://github.com/DivyanshSethi000/snn-ga-optimisation.

## IV. MOO FOR SPONTANEOUSLY ACTIVE CORTICAL RSNN

In Case 1, we ran 15 trials of simulations per parameter set and for GA, 25 generations with generation size of 50, based on their appropriate convergences (not shown). Fig. 3A shows sample simulated spike raster diagrams (neuron number vs time) of the network model with the lowest overall RMSE value for each set of targeted excitatory (exc, blue) and inhibitory (inh, green) neuronal firing rates. In both, target oscillation frequency of 50 Hz was used. Spontaneous firing with visible network oscillatory dynamics could already be observed. When excitatory neurons were set at higher firing rates than that of inhibitory neurons (Fig. 3B, left column) we could see the aggregated excitatory (blue) and inhibitory (green) neuronal population firing rates were comparable with their corresponding target values (dashed blue and green, respectively). The order of the neuronal firing rates were the same as that of the targets. The corresponding dominant network oscillation frequency was 38.8 Hz (Fig. 3C, red dashed), within the gamma band frequency, and rather close to the targeted value of 50 Hz in the same band (Fig. 3C, black dashed). The above results were obtained from the lowest overall aggregated RMSE of 0.211 (Fig. 3D, purple arrow), which were embedded within a Pareto frontier of possible optimal solutions (Fig. 3D).

We further found that GA could also nearly satisfy the condition (overall RMSE of 0.541) where the target excitatory and inhibitory neuronal firing rates were 2 Hz and 2 Hz, respectively, and oscillation frequency at 50 Hz (Fig. 3B, middle column). However, when the target excitatory and inhibitory neuronal firing rates were set at 2 Hz and 10 Hz, respectively, and target oscillation frequency at 50 Hz, GA could not satisfactorily (overall RMSE of 0.584) reach these targets (Fig. 3B, right column). This could be due to model's structural constraints, e.g. instantaneous synaptic dynamics. Interestingly, for all the sets of targets investigated, the connectivity sparsity *f* parameter attained close to 1, i.e. all-to-all connectivity, except the last evaluation. The results here provided a good starting point for further evaluations.

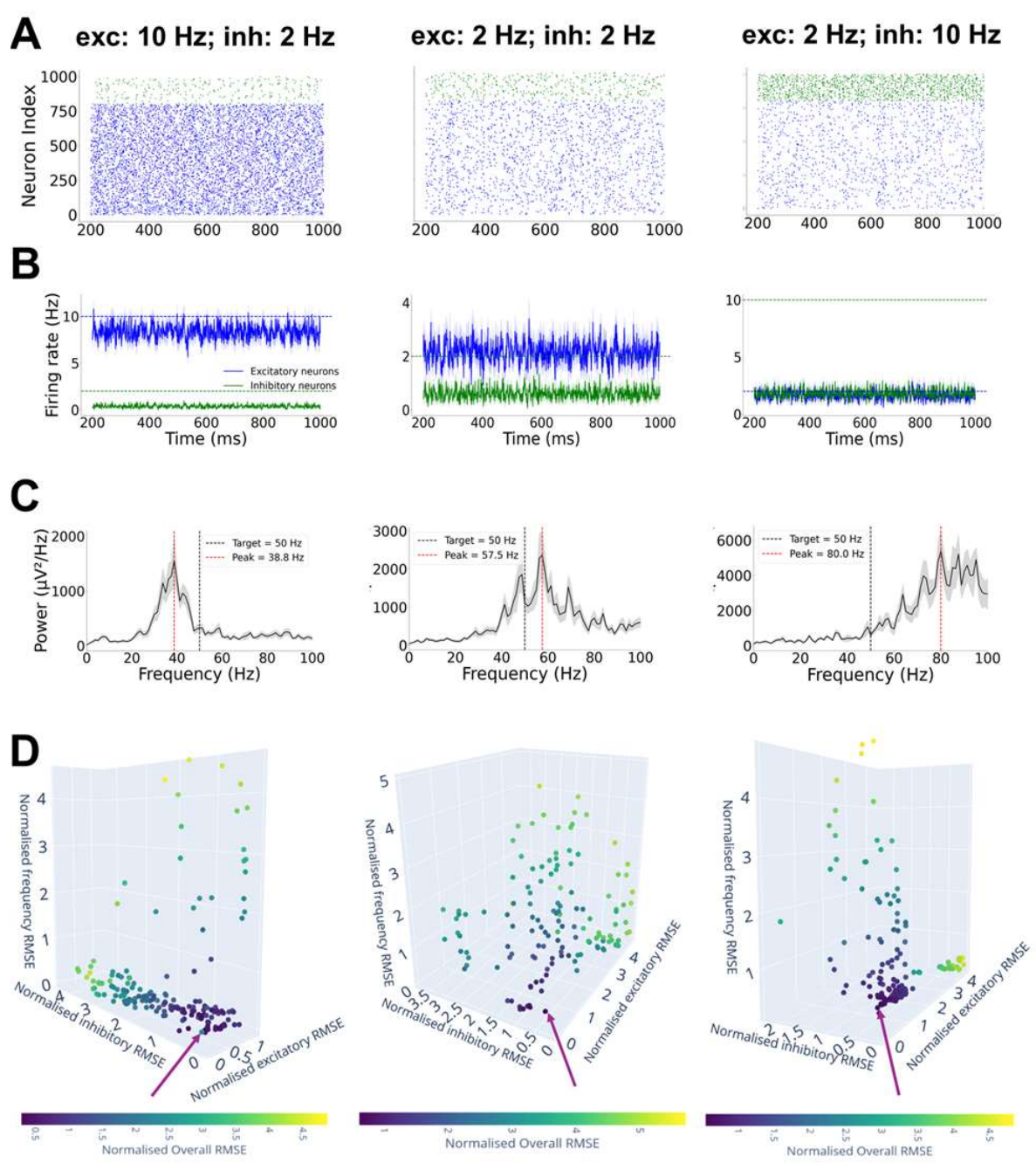

*Fig. 3. GA application on spontaneously active cortical RSNN of Izhikevich type. Left, middle and right column: optimal solutions with different set of target excitatory (exc) and inhibitory (inh) neuronal firing rates, with 50 Hz oscillation frequency. (A) Sample simulation trial of the 1000 spiking neurons over 1000 ms simulated time. Dot: spike time for a neuron. Blue (green): excitatory (inhibitory). (B) Firing rates averaged over neurons and trials with 1 ms window. Dashed lines: Targets for excitatory (10 Hz) and inhibitory (2 Hz) neurons. (C) PSD showed dominant network oscillation frequency (highest peak; red dashed) with respect to target frequency (black dashed). (D) Pareto frontier with lower overall RMSEs in darker colours. Purple arrows denote lowest overall RMSE values used in (A-C) of 0.211 (left column), 0.541 (middle column) and 0.584 (right column).*

## V. MOO FOR MODELLED BRAIN ORGANOID WITH LOW SPONTANEOUS ACTIVITIES

As in Case 1, we ran one trial of simulations per parameter set, as in the data preprocessing, and 25 generations and generation size of 50 in GA, using the 0.09 Hz of population- and time-averaged firing rate and 0.195 Hz of population dominant frequency as the objectives. Given the difficulty of attaining low neural activities, we fixed thalamic input to excitatory neurons to be 2.7 and that of inhibitory neurons to be 0.9, and set bounds for parameters *ge*, *gi* and *f* to be [0.7, 1.1], [0.8, 1,1] and [0.8, 1.0], respectively.

We found that a sample good final model parameters could readily recapitulate the averaged firing rate at 0.14 Hz (Fig. 4A), with a dominant frequency of 0.153 Hz (Fig. 4B), both very close to their targets. A sample spike rastergram of simulated trial was not shown due to too sparse spiking activity. The full Pareto frontier was also not illustrated due to highly localised optimal solution (normalised frequency RMSE ~0.095; normalised firing rate RMSE ~2.45). Hence, this indicate that GA can readily optimise small RSNN model with small number of neurons for low neuronal population firing rate and oscillation frequency in brain organoid.

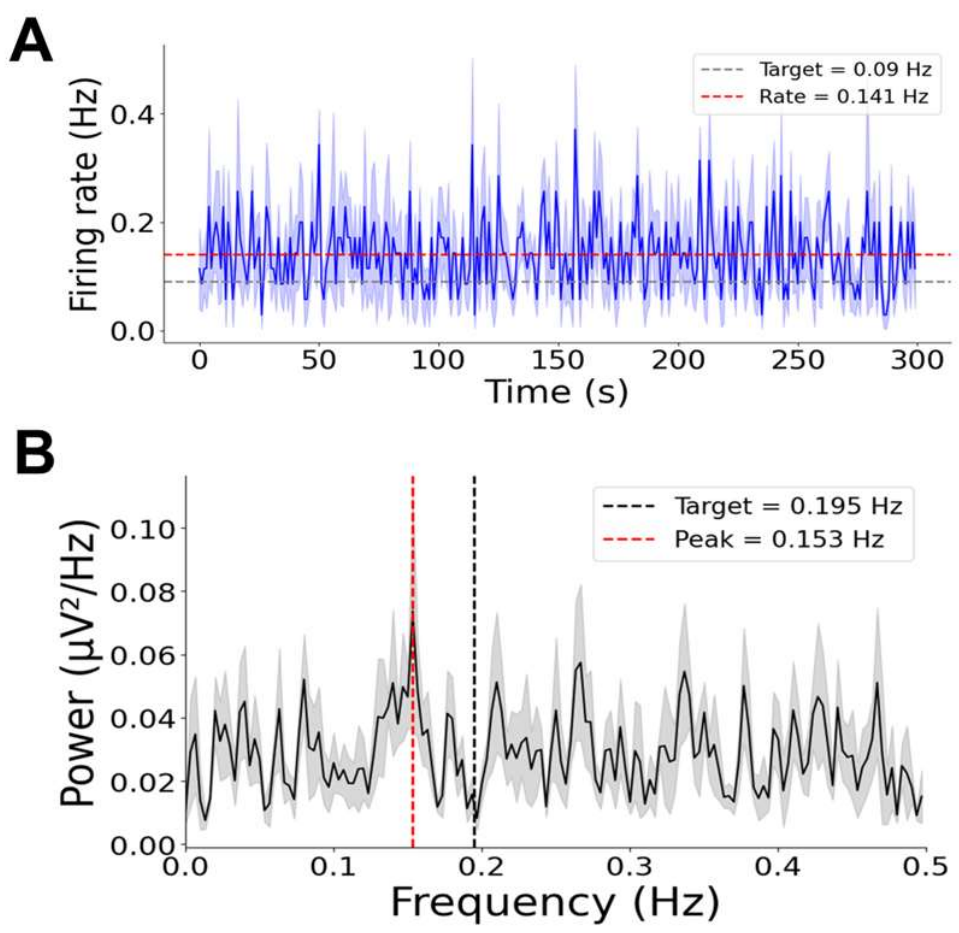

*Fig. 4. SNN modelling of brain organoid sample with low spontaneous activities and its GA application on the SNN model. Optimal solutions with normalised overall RMSE of 0.11. (A) Firing rate averaged over 5 excitatory neurons and 2 inhibitory neurons over 300 s simulated time with 1 s window. Dashed grey (red) line: Target (averaged) value. (B) PSD showed dominant network oscillation frequency (highest peak; red dashed) with respect to target frequency (black dashed). Pareto frontier not included due to highly localised optimal solution. Spike rastergram of sample simulated trial not included to sparse activity.*

## VI. MOO FOR NON-STATIONARY ACTIVITIES IN DECISION RSNN

In Case 3, we investigated how we could use GA to search for model connectivity parameters that could exhibit different conditions across two time epochs, one before stimulus onset at 500 ms (epoch I) and the other during stimulus presentation (epoch II). The challenge was that the competing excitatory neuronal population firing rates had to be small in epoch I's 200-500 ms time duration (for fair start) while exhibiting clear WTA behaviour in epoch II.

We ran 30 trials of simulations per parameter set and for GA, 20 generations with generation size of 36, based on their appropriate convergences (not shown). Fig. 5A shows a sample simulated spike raster diagram (neuron number vs time) of the network model at the 20$^{th}$ generation. The model exhibited common excitatory neuronal firing rate levels prior to stimulus onset time in epoch I. Upon stimulus presence in epoch II, one excitatory neuronal population's activity (blue – encoding choice 1) was enhanced at the expense of the other (encoding choice 2) – WTA behaviour for making a unitary choice. Moreover, the network oscillatory dynamics became more visible in epoch II, especially for the inhibitory population (green) and the winning population (blue).

In Fig. 5B, the subpopulation firing rates more clearly exhibited WTA behaviour for making two difference choices (superimposed). The targeted firing level criteria were also well satisfied. The ramping up of the winning population activities was consistent with the evidence accumulation framework for decision formation [36][37]. Even without optimising for epoch III, we found temporary storage of the choice made through decaying persistent firing rates [37] that could last for as long as 500 ms.

The PSDs averaged over both choices (Fig. 5B, insets) across time epochs, showed that during stimulus presence, the dominant network frequency (22 Hz) reached rather closely to its targeted value of 30 Hz, hence simultaneously satisfying another criterion. The maximal peak frequencies before and after stimulus presentation were lower than 20 Hz, although double peaks could also be observed. The overall possible solutions were represented as Pareto frontier in Fig. 5C.

## VII. Conclusion and discusson

In this work, we have successfully applied a version of the GA algorithm, called NSGA-III, in search for optimal model parameter of a spontaneously active canonical cortical column RSNN model, a low spontaneously active brain organoid via RSNN modelling, and a low-firing cortical RSNN model that exhibits decision-making behaviour. Importantly, we were able to identify sets of optimal connectivity parameters, within Pareto frontiers, with respect to neuronal population firing rates and network oscillatory dynamics. This was evaluated using RMSEs of these features with respect to their targeted values.

Our work was an extension of our previous MOO work [32] that optimised only for neuronal population firing rates in the cortical RSNN, similar to our Case 1. We extended to include network oscillatory dynamics, actual biological brain organoid, and a functional RSNN for making binary decisions with transient neural dynamics. For each case and condition, we also identified the best solution within the Pareto frontiers, that is, with the lowest overall RMSE.

Despite the progress made, there were limitations in our study. With additional sensitivity tests performed, we found excitatory neuronal firing rates, including their WTA separability for decision-making, to emerge as the more flexible variables in meeting objectives, while inhibition and oscillatory frequency remained more structurally constrained. Particularly, spontaneous canonical cortical oscillations could shift more flexibly to lower than higher frequencies, while decision-phase oscillations were even further constrained. These results reveal clear flexibility limits in how each SNN can adjust its firing-rate and oscillatory dynamics. Future work will study RSNN model structure-function relationship. For the brain organoid, we had only applied to spontaneous activity. In future work, our "digital twinning" approach here could be extended to brain organoid computing [38][39], to provide insights into the neural circuit mechanisms underlying their computation. We could also optimise for single neuronal conditions rather than averaging over all neurons. However, more electrophysiological and -omics information would be required to identify specific neuronal types to model. For the decision RSNN, we had currently optimised for firing rates and network oscillation frequencies during unbiased stimulus presence. Future work could involve oscillation frequencies in several time epochs, with stimulus biases and choice behavioural measures (e.g. decision latency and choice accuracy) [16][36][37].

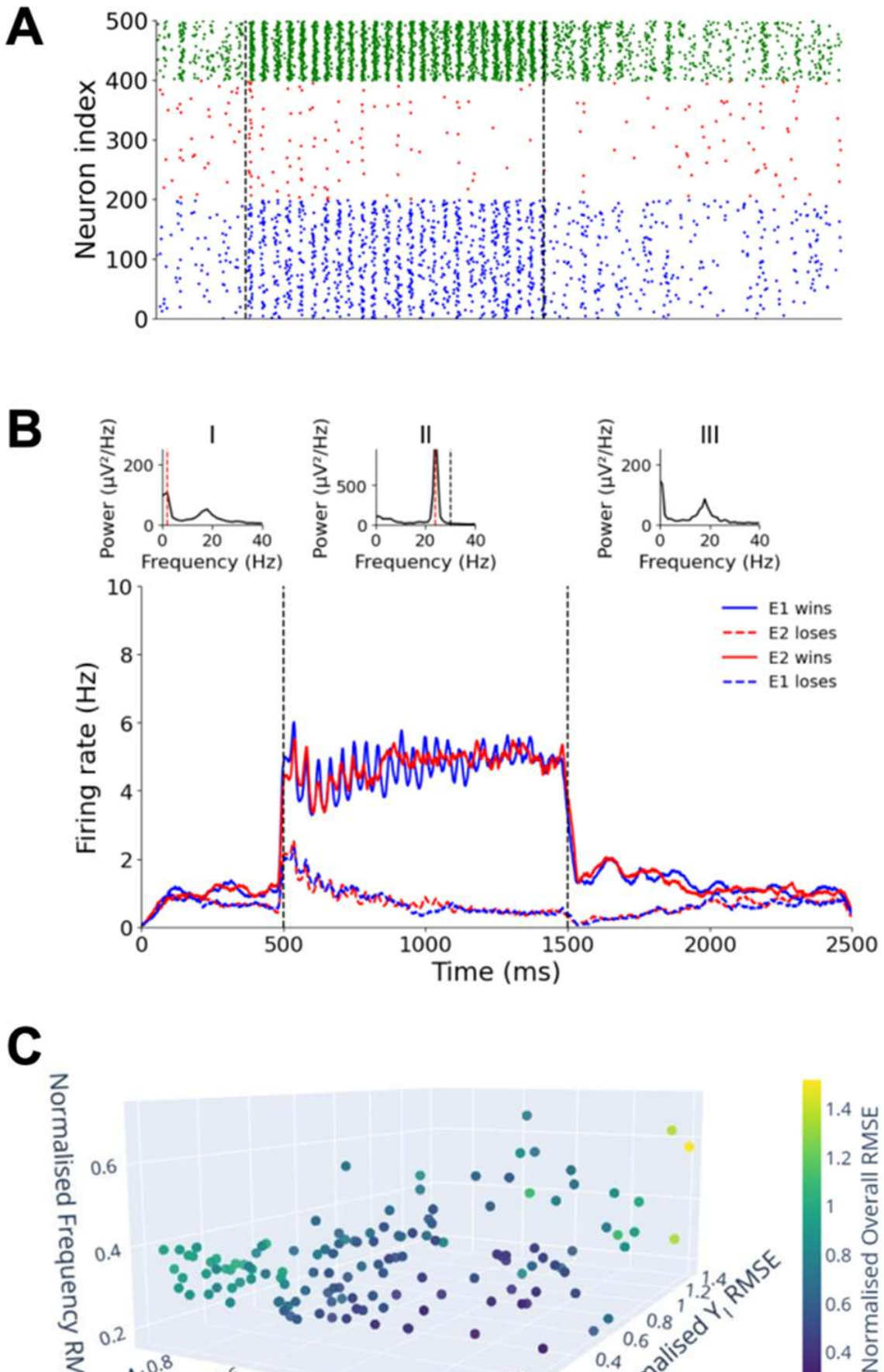

*Fig.5. GA application on RSNN model for binary decision-making with unbiased stimulus inputs. (A-B) Optimised results at the 20$^{th}$ generation, with normalised overall RMSE 0.20. (A) Sample simulation trial of 500 neurons over 2500 ms simulated time. Dot: spike time for some neuron. 400 excitatory neurons, equally divided into 2 competing competing populations (blue and red), and 100 inhibitory neurons (green). Dashed lines: stimulus onset and offset. (B) Competing population firing rates averaged over neurons and trials with 1 ms window. Data sorted into two different choices represented by winning blue or red. Bold (dashed) red/blue: winning (losing) activities. Horizontal dashed lines: criteria for winner (high) and loser (low). Insets: PSDs showed dominant network oscillation frequencies in each time epoch; in epoch II, frequency with highest peak (red dashed) were compared to*

*target frequency (green dashed). (C) Pareto frontier with lower overall RMSEs as darker dots.* $Y_k = |f_{E1,k} - f_{E2,k}|$ *denotes the difference of the competing firing rates,* $f_{E1}$ *and* $f_{E2}$*, in epoch* $k = I$ *or II.*

## ACKNOWLEDGMENT

We thank FinalSpark for accessing its Neuroplatform data, and Saugat Bhattacharyya and Brendan Lenfesty for helpful discussions. This work was supported by HSC R&D (STL/5540/19) and MRC (MC_PC_20020). We are grateful for access to the Tier 2 High-Performance Computing resources provided by the NI-HPC facility funded by the UK Engineering and Physical Sciences Research Council (EPSRC), Grant No. EP/T022175/1.

## REFERENCES

[1] A. Taherkhani, A. Belatreche, Y. Li, G. Cosma, L. P. Maguire and T. M. McGinnity, “A review of learning in biologically plausible spiking neural networks,” Neural Netw., vol. 122, pp. 253–272.

[2] W. Maass, “Networks of spiking neurons: the third generation of neural network models,” Neural Netw., vol. 10, pp. 1659–1671, 1997.

[3] H. Paugam-Moisy and S. M. Bohte, “Computing with Spiking Neuron Networks,” In: Rozenberg, G., Bäck, T., Kok, J.N. (eds) Handbook of Natural Computing. Springer, Berlin, Heidelberg, pp. 335-376,. https://doi.org/10.1007/978-3-540-92910-9_10.

[4] F. Ponulak and A. Kasinski, “Introduction to spiking neural networks: Information processing, learning and applications,” Acta Neurobiol. Exp., vol. 71, pp. 409–433, 2011.

[5] N. K. Kasabov, “Spiking neural networks for deep learning and knowledge representation,” Neural Netw., vol. 119, pp. 341-342, 2019, doi: 10.1016/j.neunet.2019.08.019.

[6] S. B. Furber, D. R. Lester, L. A. Plana, J. D. Garside, E. Painkras, S. Temple and A. D. Brown., “Overview of the SpiNNaker System Architecture,” IEEE Trans. Comput., vol. 62, pp. 2454– 2467, 2013.

[7] J. Pei, L. Deng, S. Song, M. Zhao, Y. Zhang, S. Wu, G. Wang, Z. Zou, Z. Wu, W. He, F. Chen, N. Deng, S. Wu, Y. Wang, Y. Wu, Z. Yang, C. Ma, G. Li, W. Han, H. Li, H. Wu, R. Zhao, Y. Xie and L. Shi, “Towards artificial general intelligence with hybrid Tianjic chip architecture,” Nature, vol. 572, pp. 106–111, 2019.

[8] L. F. Abbott, and P. Dayan, *Theoretical Neuroscience*. MIT Press, Cambridge, MA, USA, 2001.

[9] W. Gerstner, W. M. Kistler, R. Naud and L. Paninski, *Neuronal Dynamics: From Single Neurons to Networks and Models of Cognition*. Cambridge: Cambridge University Press, Cambridge, 2014.

[10] E. M. Izhikevich, *Dynamical Systems in Neuroscience: The Geometry of Excitability and Bursting*. The MIT Press, Cambridge, MA USA, 2007.

[11] M. Taketani and M. Baudry, *Advances in Network Electrophysiology: Using Multi-Electrode Arrays*. Springer, Singapore, 2006.

[12] D. L. Schomer and F. H. Lopes da Silva, Niedermeyer's Electroencephalography: Basic Principles, Clinical Applications, and Related Fields. 7th Ed., Oxford University Press, 2017.

[13] H. Op de Beeck and C. Nakatani, *Introduction to Human Neuroimaging*. 2nd Ed., Cambridge University Press, Cambridge, 2025.

[14] W. Singer and F. Effenberger, “Oscillations in natural neuronal networks; an epiphenomenon or a fundamental computational mechanism?” Hu Arenas, vol. 8, pp. 846–868, 2025.

[15] X.-J. Wang, “Neurophysiological and computational principles of cortical rhythms in cognition,” Physiol. Rev., vol. 90, pp. 1195–1268, 2010.

[16] A. Azimi and K. Wong-Lin, “Neural oscillation as a selective modulatory mechanism on decision confidence, speed, and accuracy,” J. Neurosci., vol. 45, e0880252025. doi: 10.1523/JNEUROSCI.0880-25.2025.

[17] E. M. Izhikevich, “Simple model of spiking neurons,” IEEE Trans. Neural Netw., vol. 14, no. 6, pp. 1569–1572, 2003. doi: 10.1109/TNN.2003.820440.

[18] E. M. Izhikevich, “Which neuron model to use?,” IEEE Trans. Neural Netw., vol. 15, pp. 1036–1070, 2004.

[19] S. Furber and S. Temple, “Neural systems engineering,” J. R. Soc. Interface, vol. 4 (13), pp. 193–206, 2007.

[20] A. Cassidy and A. G. Andreou, "Dynamical digital silicon neurons," In: *2008 IEEE Biomedical Circuits and Systems Conference*, Baltimore, MD, USA, 2008, pp. 289-292, doi: 10.1109/BIOCAS.2008.4696931.

[21] W. Yan and J. Qiu, “Neuromorphic computing in sensory systems: A review,” J. Neuro. Intel., vol. 1, pp. 9-20, 2024.

[22] S. Kampakis, “Improved Izhikevich neurons for spiking neural networks,” J. Soft Comput., vol. 16(6), pp. 943-953, 2012. DOI: 10.1007/s00500-011-0793-1.

[23] G. E. Soares, H.E. Borges, R. M. Gomes, G. M. Zeferino and A. P. Braga, “Emergence of synchronicity in a self-organizing spiking neuron network: an approach via genetic algorithms,” Nat. Comput., vol. 11, pp. 405–413, 2012.

[24] K. Wong-Lin, A. Joshi, G. Prasad and T. M. McGinnity, “Network properties of a computational model of the dorsal raphe nucleus,” Neural Netw. Vol. 32, pp. 15–25, 2012.

[25] X. Fang, S. Duan and L. Wang, “Memristive Izhikevich spiking neuron model and its application in oscillatory associative memory,” Front. Neurosci., vol. 16, 885322, 2022. doi: 10.3389/fnins.2022.885322.

[26] S. Haykin, *Neural Networks and Learning Machines*. 3rd Ed., Pearson Education, Inc., McMaster University, Hamilton, 2009.

[27] M. J. Kochenderferm and T. A. Wheeler, *Algorithms for Optimization*. The MIT Press, Cambridge, MA, USA, 2019.

[28] M. Negnevitsky, *Artificial Intelligence: A Guide to Intelligent Systems*. 4th Ed., Pearson Education, 2024.

[29] M. Mitchell, *An Introduction to Genetic Algorithms*. MIT Press, Cambridge, MA, USA, 2002.

[30] K. Deb, “Multi-objective Optimisation Using Evolutionary Algorithms: An Introduction.” In: Wang, L., Ng, A., Deb, K. (eds) Multi-objective Evolutionary Optimisation for Product Design and Manufacturing. Springer, London, 2011. https://doi.org/10.1007/978-0-85729-652-8_1

[31] I. Ezenwe, A. Joshi and K. Wong-Lin, "Genetic algorithmic parameter optimisation of a recurrent spiking neural network model," *2020 31st Irish Signals and Systems Conference (ISSC)*, Letterkenny, Ireland, 2020, pp. 1-6, doi: 10.1109/ISSC49989.2020.9180185.

[32] J. Fitzgerald, and K. Wong-Lin, “Multi-objective optimisation of cortical spiking neural networks with genetic algorithms,” *2021 32nd Irish Signals and Systems Conference (ISSC)*, Athlone, Ireland, 2021, pp. 1-6, doi: 10.1109/ISSC52156.2021.9467860.

[33] K. Deb and H. Jain, “An Evolutionary Many-Objective Optimization Algorithm Using Reference-Point-Based Nondominated Sorting Approach, Part I: Solving Problems With Box Constraints,” vol. 18, no. 4, pp. 577–601, 2014.

[34] L. D. R. Oliveira, R. M. Gomes, B. A. Santos and H. E. Borges, “Effects of the parameters on the oscillation frequency of Izhikevich spiking neural networks,” Neurocomputing, vol. 337, pp. 251-261, 2019.

[35] K. Tasnim and J. Liu, “Emerging bioelectronics for brain organoid electrophysiology,” J. Mol. Biol., vol. 434(3), 167165, 2022. doi: 10.1016/j.jmb.2021.167165.

[36] X.-J. Wang, “Probabilistic decision making by slow reverberation in cortical circuits,” Neuron, vol. 36(5), pp. 955-968, 2002.

[37] K.-F Wong and X.-J. Wang, “A recurrent network mechanism of time integration in perceputal decisions,” J. Neurosci., vol. 26(4), pp. 1314-1328, 2006.

[38] F. D. Jordan, M. Kutter, J.-M. Comby, F. Brozzi and E. Kurtys, “Open and remotely accessible Neuroplatform for research in wetware computing,” Front. Artif. Intell., vol. 7, 1376042, 2024. doi: 10.3389/frai.2024.1376042.

[39] L. Smirnova, B. S. Caffo, D. H. Gracias, Q. Huang, I. E. Morales Pantoja, B. Tang, D. J. Zack, C. A. Berlinicke, J. L. Boyd, T. D. Harris, E. C. Johnson, B. J. Kagan, J. Kahn, A. R. Muotri, B. L. Paulhamus, J. C. Schwamborn, J. Plotkin, A. S. Szalay, J. T. Vogelstein, P. F. Worley and T. Hartung, “Organoid intelligence (OI): the new frontier in biocomputing and intelligence-in-a-dish,” Front. Sci., vol. 1, 1017235, 2023. doi: 10.3389/fsci.2023.1017235D.